\documentclass[aps,pra,floatfix,twocolumn,floatfix]{revtex4}
\usepackage{dcolumn,bm,graphicx,amsmath}

\begin{document}

\title{Blackbody radiation
shift for the $^1$S$_0$ - $^3$P$_0$ optical clock transition in zinc and cadmium atoms}
\author{Vladimir A. Dzuba}
\affiliation{School of Physics, University of New South Wales, Sydney 2052,
Australia}
\affiliation{Department of Physics, University of Nevada, Reno, Nevada 89557}
\author{Andrei Derevianko}
\affiliation{Department of Physics, University of Nevada, Reno, Nevada 89557}

\date{\today}

\begin{abstract}
Black-body radiation (BBR)  shifts of $^3\!P_0-^1\!S_0$ clock
transition in divalent atoms Cd and Zn are evaluated  using accurate relativistic
many-body techniques of atomic structure. Static polarizabilities of the clock levels
and relevant electric-dipole matrix elements
are  computed. We also present a comparative overview of the BBR shifts
in optical clocks based on neutral divalent atoms trapped in optical lattices.
\end{abstract}

% 2006 PACS
%06.30.Ft Time and frequency
%32.10.Dk Electric and magnetic moments, polarizability
%31.25.-v Electron correlation calculations for atoms and molecules

\maketitle

%\section{Introduction}

One of the factors limiting the accuracy of the modern atomic clocks is the perturbation of the clock frequency
by the bath of thermal photons, i.e., by black body radiation (BBR).  $10^{-15}$ is the typical value of the fractional BBR correction to optical lattice clocks~\cite{PorDer06BBR} at room temperatures, while the current generation of optical atomic clocks have demonstrated the fractional inaccuracies at the level of $10^{-18}$ or better~\cite{Brewer2019,Bothwell2019,McGrew2018}. Therefore, all the recent advances in atomic clocks address the BBR shift problem either through cryogenic techniques, active temperature stabilization, or  specially-designed BBR chambers. All of these techniques can be advanced further by using atoms that have a reduced sensitivity to BBR.   

%Due to its sizable effect on the output of high-accuracy clocks, the black-body-radiation (BBR) shift has to be 
%explicitly enters the present definition of the
%unit of time, the second. The output of the primary Cs frequency standard has to be adjusted using the standardized value of
%the BBR coefficient. 
To the leading order, the fractional BBR correction to the unperturbed clock frequency $\nu_0$ can be parameterized as
\[
  \frac{\delta \nu}{\nu_0} = \beta \left(\frac{T}{300 K} \right)^4 \, ,
\]
where $T$ is the bath temperature.
There are two issues associated with the BBR shift: (i) one needs to know the coefficient $\beta$ with sufficiently high accuracy so that the uncertainty in $\beta$ does not degrade the clock output and (ii) even if $\beta$ is known precisely, there are uncertainties arising from the
ambient temperature fluctuations and imperfect knowledge of the temperature field. 
Apparently, the smaller the $\beta$, the better.

There are two main classes of optical atomic clocks that are presently well-positioned to eventually replace the primary frequency
standard. The first, more mature, class of clocks is based on trapped ions and the second class employs neutral divalent atoms trapped in
optical lattices. A comparative overview of the BBR shift for various ion clocks is given in Ref.\cite{RosItaSch06} and for lattice clocks in Ref.~\cite{PorDer06BBR}. The NIST group~\cite{RosSchHum07} has pointed out that the BBR shift is exceptionally small in Al$^{+}$ ion, $\delta{\nu}/{\nu_0} \sim 10^{-17}$.
For divalent atoms  considered in the literature so far (Mg, Ca, Sr, Yb, Hg) the least susceptible are mercury lattice clocks~\cite{HacMiyPor08}, $\delta{\nu}/{\nu_0} \sim 10^{-16}$ at room temperatures. 

Divalent cadmium and zinc atoms were found recently~\cite{OvsPalTai07} to have properties suitable
for realizing the neutral atom optical lattice clocks.
With the BBR shift being one of the most important contributors to the uncertainty budget of the clocks,
here we extend the survey of Ref.~\cite{PorDer06BBR} and compute the BBR shifts for the Cd and Zn lattice clocks.
The results of our analysis are summarized in Table~\ref{Tab:BBRsum}.
We find that for Cd and Zn the fractional BBR shifts are comparable to the so-far most favorable Hg.
At least from this perspective, these atoms may serve
as a competitive alternative to already operational Sr, Yb, and Hg clocks.

\begin{table}[h]
\caption{Black-body radiation shift for the clock transitions between the
$^1$S$_0$ ground state and the
lowest-energy $^3$P$^o_0$  state for divalent atoms.
Result for Zn and Cd are from this work, for Mg, Ca, Sr, Yb from Ref.~\cite{PorDer06BBR}, and for
Hg from Ref.~\cite{HacMiyPor08}.
$\protect%
\delta\protect\nu$ is the BBR shift at $T=300\,\mathrm{K}$
with our estimated uncertainties. $\protect\nu_0$ is the clock transition
frequency, and $\protect\delta\protect\nu/\protect\nu_0$ is
the fractional contribution of the BBR shift. The last column lists
fractional errors in the absolute transition frequencies induced by the
uncertainties in the BBR shift. }
\label{Tab:BBRsum}%
\begin{tabular}{lcccc}
\hline \hline
Atom
&\multicolumn{1}{c}{$\delta\nu$, (Hz)}
&\multicolumn{1}{c}{$\nu_0$, (Hz)}
&\multicolumn{1}{c}{$\delta\nu/\nu_0$}
&\multicolumn{1}{c}{Uncertainty}\\
\hline
Zn  &  -0.244(10) & $9.69 \times 10^{14}$ & $-2.5 \times 10^{-16}$ & $1  \times 10^{-17}$\\
Cd  &  -0.248(15) & $9.03 \times 10^{14}$ & $-2.8 \times 10^{-16}$ & $2 \times 10^{-17}$\\
\hline
Mg~\cite{PorDer06BBR}   &  -0.258(7)  & $6.55 \times 10^{14}$ & $-3.9  \times 10^{-16}$ & $1  \times 10^{-17}$\\
Ca~\cite{PorDer06BBR}   &  -1.171(17) & $4.54 \times 10^{14}$ & $-2.6  \times 10^{-15}$ & $4  \times 10^{-17}$\\
Sr~\cite{PorDer06BBR}   &  -2.354(32) & $4.29 \times 10^{14}$ & $-5.5  \times 10^{-15}$ & $7  \times 10^{-17}$\\
Yb~\cite{PorDer06BBR}   &  -1.25(13)  & $5.18 \times 10^{14}$ & $-2.4  \times 10^{-15}$ & $3  \times 10^{-16}$ \\
Hg~\cite{HacMiyPor08}   &   -0.181    & $1.13 \times 10^{15}$ & $-1.6  \times 10^{-16}$ &     \\
\hline \hline
\end{tabular}
\end{table}

%In addition to computing the BBR shifts on the clock transition for Cd and Zn we also determine the magic frequencies for operating
%trapping lattice lasers. At the magic frequency both clock levels are shifted by the laser field by the very same amount -- the clock frequency
%remains unperturbed by the trapping~\cite{YeKimKat08}. While these magic wavelength were determined earlierF=m

{\em Details of calculations ---}
To compute the energy shift due to black-body radiation we use the
formalism developed in Ref.~\cite{PorDer06BBR}. The electric-dipole contribution to
the BBR energy shift of state $v$ is given by
\begin{eqnarray}
&&\delta E_{v} \approx -\frac{2}{15} (\alpha \pi)^3 T^4
\alpha_v(0) \left[ 1 + \eta \right] \, ,  \label{Eq:delEg} \\
&&\eta = \frac{(80/63) \pi^2}{\alpha_v(0) T} \sum_p \frac{ |\langle
p||D||v \rangle|^2 }{(2J_v+1) y_p^3}\left( 1+ \frac{21
\pi^2}{5y_p^2} + \frac{336 \pi^4}{11y_p^4} \right) \, .  \notag
\end{eqnarray}
Here $y_p = (E_{p}-E_v)/T$, $\alpha_v(0)$ is the
static scalar dipole polarizability, and $\eta$ represents a ``dynamic'' fractional correction to the total shift. $D$ is the electric-dipole
operator.
The calculations requires evaluating the static polarizability for both clock levels.
The clock transition is between the $^1$S$_0$ ground state and the
lowest-energy $^3$P$^o_0$  state.

%To arrive at the above equation, we used

%asymptotic expansion $F_{1}\left( y\right) \approx \frac{4\pi ^{3}}{45y}+
%\frac{32\pi ^{5}}{189y^{3}}+ \frac{32\pi ^{7}}{45y^{5}}+\frac{512\pi ^{9}}{
%99y^{7}}$, which has an accuracy better than 0.1\% for $|y|>10$.

The static scalar polarizability $\alpha_v(0)$ of an atom in state $v$ is given by
\begin{equation}
  \alpha_v(0) = \frac{2}{3(2J_v+1)} \sum_n \frac{|\langle v||\mathbf{D}
|| n \rangle|^2}{E_n-E_v},
\label{eq:alpha}
\end{equation}
where
summation goes over the complete set of excited many-body states (including continuum and core-excited states).
We use the Dalgarno-Lewis method and  reduce the summation to
solving  the inhomogeneous Schr\"{o}dinger (Dirac) equation (setup is similar to
Ref.~\cite{DerJoh97}). In this approach, a correction to the atomic wave
function due to the external electric field is introduced
\begin{equation}
  \langle \delta \Psi_v |= \sum_n \frac{\langle v||\mathbf{D}|| n \rangle}{E_n-E_v}
  \langle n|.
\label{eq:deltaPsi}
\end{equation}
This correction satisfies an inhomogeneous equation
\begin{equation}
  (\hat H_0 - E_v) \delta \Psi_v = - \mathbf{D} \Psi_v,
\label{eq:H0deltaPsi}
\end{equation}
where $\hat H_0$ is an effective Hamiltonian of the atom. Once the $\delta
\Psi_v$ is found, static polarizability is calculated as
\begin{equation}
  \alpha_v(0) = \frac{2}{3(2J_v+1)}\langle \delta \Psi_v|| \mathbf{D}
  ||v\rangle .
\label{eq:alpha-deltaPsi}
\end{equation}

We employ a computational scheme based on combining the configuration interaction method with the many-body perturbation
theory (CI+MBPT)~\cite{DzuFlaKoz96}. The
effective Hamiltonian is constructed for the two valence electrons, while
excitations from the core are taken into account by means of the MBPT.
The Hamiltonian has the form
\begin{equation}
  \hat H_0 = \hat h_1(1) + \hat h_1(2) + \hat h_{12},
\label{eq:H0}
\end{equation}
where $\hat h_1$ is a single-electron part of the relativistic Hamiltonian
\begin{equation}
  \hat h_1 = c(\mathbf{\alpha \cdot p}) + mc^2(\beta-1) - \frac{Ze^2}{r}
  + \hat V_{\rm core} + \hat \Sigma_1 \, .
\label{eq:h1}
\end{equation}
Here $c$ is speed of light, and $\mathbf{\alpha}$ and $\beta$ are Dirac matrices,
$Ze$ is the nuclear charge, $\hat V_{\rm core}$ is the Hartree-Fock potential of
the atomic core (including the non-local exchange term) and $\hat \Sigma_1$ is
the correlation potential which describes the correlation interaction between
a valence electron and the core (see Refs.\cite{DzuFlaKoz96,DzuJoh98} for details).

The $\hat h_{12}$ operator in (\ref{eq:H0}) is the two-electron part of the
Hamiltonian:
\begin{equation}
  \hat h_{12} = \frac{e^2}{r_{12}} + \hat \Sigma_2,
\label{eq:h12}
\end{equation}
where first term is standard Coulomb interaction between valence electrons and
second term is the correction to it due to correlations with core electrons.

We use the second-order MBPT to calculate the self-energy operators
$\hat \Sigma_1$ and $\hat \Sigma_2$
via direct summation over a complete set of single-electron states. This set of
basis states is constructed using the B-spline technique~\cite{JohSap86}. We
use 40 B-splines of order 9 in a cavity of 40 Bohr radius. The same basis of the
single-electron states is also used in constructing the two-electron basis states for
the CI calculations. We employ partial waves $\ell=0-4$ for the valence CI subspace and
$\ell=0-5$ for internal summations inside the self-energy operator.

%To estimate the uncertainty of the calculations we perform all calculations in
%two different variants. One is pure {\em ab initio} and in the second variant
Additionally, to mimic the omitted higher-order MBPT effects,
we rescale the $\hat \Sigma$ operator to fit the experimental energies. The $\hat
\Sigma_1$ operator is replaced in (\ref{eq:H0}) by $\lambda_l \hat \Sigma_1$,
where $l=0,1,2$ is a the angular momentum of a single-electron state. The
$\hat \Sigma_2$ operator is replaced by $f_k \hat \Sigma_2$, where $k$ is
multipolarity of the Coulomb interaction. The values of the rescaling
parameters are presented in Table~\ref{Tab:rescaling}.
The resulting energies after the scaling procedure are listed in Table~\ref{Tab:Energies}.
A typical deviation from the experimental values is in the order of 100 $1/\mathrm{cm}$. Even after
the scaling, the disagreement remains, as the number of fitting parameters is limited.

\begin{table}[h]
\caption{Rescaling parameters $\lambda_l$ and $f_k$ for the core-valence
  correlation operators $\hat \Sigma_1$ and $\hat \Sigma_2$}
\label{Tab:rescaling}
\begin{tabular}{cccc}
\hline \hline
& $\lambda_s$ & $\lambda_p$  & $f_1$ \\
\hline
Zn & 1.113  & 1.106  & 1 \\
Cd & 0.8714 & 0.887  & 0.8 \\
\hline \hline
\end{tabular}
\end{table}

\begin{table}[h]
\caption{Energy levels of Zn and Cd (cm$^{-1}$); comparison of ``scaled'' theory and
  experiment. $n=4$ for Zn and $n=5$ for Cd.}
\label{Tab:Energies}
\begin{tabular}{llccccc}
\hline \hline
&&&\multicolumn{2}{c}{Zn} &\multicolumn{2}{c}{Cd} \\
Config.    & State     &$J$& Expt.  & Theory & Expt.  & Theory \\
\hline
$nsnp$     & $^3$P$^o$ & 0 &  32311 &  32348 &  30114 &  30108 \\
           &           & 1 &  32501 &  32546 &  30656 &  30664 \\
           &           & 2 &  32890 &  32950 &  31827 &  31866 \\

$nsnp$     & $^1$P$^o$ & 1 &  46745 &  46908 &  43692 &  43721 \\

$ns(n+1)s$ & $^3$S     & 1 &  53672 &  53412 &  51484 &  51317 \\

$ns(n+1)s$ & $^1$S     & 0 &  55789 &  55513 &  53310 &  53088 \\

$nsnd$     & $^1$D     & 2 &  62459 &  62333 &  59220 &  59282 \\

$nsnd$     & $^3$D     & 1 &  62769 &  62606 &  59486 &  59512 \\
           &           & 2 &  62772 &  62609 &  59498 &  59521 \\
           &           & 3 &  62777 &  62613 &  59516 &  59534 \\
\hline \hline
\end{tabular}
\end{table}

{\em Results --- }
With the computed wavefunctions, we may evaluate various matrix elements. While computing  matrix elements (and polarizabilities) we use single-particle matrix elements dressed in the
random-phase approximation. Qualitatively this correspond to the shielding of the applied electromagnetic field
by the core electrons. Notice that
the static polarizability depends sensitively on the values of the dipole matrix elements for the lowest-energy excitations.
Our computed  dipole matrix elements for the two lowest-energy excitations originating
from the two clock states are presented in Table~\ref{Tab:E1}. The inter-combination
transition $^1$S$_0 \rightarrow ^3$P$^o_1$ is non-relativistically forbidden.
A three-fold increase in the matrix element values when progressing from Zn $Z=30$ to heavier Cd $Z=48$ is consistent
with the relevant suppression factor of $(\alpha Z)^2$.
Similarly to the case of Sr and Yb atoms~\cite{YasKisTak06,TakKomHon04}, we anticipate that the high-accuracy values for the $^1$S$_0 \rightarrow ^1$P$^o_1$
matrix elements may be derived from photoassociation spectroscopy with ultracold atoms.
If such data become available, the accuracy of our
values for polarizability may be improved by correcting the matrix elements of Table~\ref{Tab:Energies} with the experimental values
and correcting $\alpha_v(0)$ with Eq.(\ref{eq:alpha}).

\begin{table}
\caption{Electric  dipole transition amplitudes (reduced matrix elements, a.u.)
 for Zn and Cd.}
\label{Tab:E1}
\begin{tabular}{ccc}
\hline \hline
Transition                      &  Zn   &  Cd   \\
\hline
$^1$S$_0 \rightarrow ^3$P$^o_1$ & 0.045 & 0.158 \\
$^1$S$_0 \rightarrow ^1$P$^o_1$ & 3.320 & 3.435 \\
$^3$P$^o_0 \rightarrow ^3$S$_1$ & 1.466 & 1.486 \\
$^3$P$^o_0 \rightarrow ^3$D$_1$ & 2.127 & 2.222 \\
\hline \hline
\end{tabular}
\end{table}

The computed values of the  static polarizabilities of the clock levels are presented in
Table~\ref{Tab:Pol}. The values combine both valence and core polarizabilities. Core polarizabilities are 2.296 a.u. for Zn and 4.971 a.u. for Cd~\cite{JohKolHua83}.
For the ground states we compare our values with the experimental results\cite{GoeHohMar96,GoeHoh95}.
For Zn our computed value is within the experimental uncertainty while for Cd the results disagree by about $2 \, \sigma$
of the experiment. Our results are consistent with the previous theoretical work~\cite{YeWan08,EllMerRer01}.
These authors employed methods sufficiently different from our approach to warrant additional confidence in the theoretical predictions.
\citet{YeWan08} used a semi-empirical model potential method and \citet{EllMerRer01} employed a
multi-reference configuration-interaction method using a two-electron relativistic pseudo-potential.

Finally, we combine the static polarizabilities using Eq.~(\ref{Eq:delEg}) and arrive
at the  BBR shifts summarized in Table~\ref{Tab:BBRsum}.
The results also include the dynamic correction $\eta$; it turns out to be less than  $7 \times 10^{-4}$ for
both Zn and Cd. This small correction can be safely neglected at the present level of accuracy.
We also estimate the theoretical error bar for the BBR correction: 4\% for Zn and 6\% for heavier Cd.
The error was evaluated by carrying out two calculations: with  and without scaling of self-energy operator
to experimental energies. We find that the resulting uncertainty would affect the accuracy of the clock output
in the 17th significant figure. Overall fractional BBR shift for both  Cd and Zn is slightly larger than in Hg, but
5 times smaller than in Sr  and 10 times smaller than in Yb.

\begin{table}[h]
\caption{Static polarizabilities of the $^1$S$_0$ and $^3$P$^o_0$ states of Zn and Cd
  (a.u.); comparison with experimental results and other calculations.}
\label{Tab:Pol}
\begin{tabular}{llllccc}
\hline \hline
Atom  & State    & Expt.~\protect\cite{GoeHohMar96,GoeHoh95} & This work &
Ref.~\cite{YeWan08}  & Ref.~\cite{EllMerRer01} \\
\hline
Zn & $^1$S$_0$           &  38.8(8)     &  38.58 &  38.12 &  39.13 \\

Zn & $^3$P$^o_0$         &              &  66.53 &  67.69 &  66.50 \\

Cd & $^1$S$_0$           &  49.65(1.62) &  46.52 &  44.63 &        \\

Cd & $^3$P$^o_0$         &              &  75.31 &  75.29 &        \\
\hline \hline
\end{tabular}
\end{table}

\begin{acknowledgments}
We would like to thank Kurt Gibble for motivating discussions.
This work was supported in part by the U.S. National Science Foundation and by the Australian
Research Council.
\end{acknowledgments}

%\bibliography{library-apd,all}

\end{document}